\documentstyle[twoside,fleqn,amssymb,epsf,espcrc2]{article}%
\begin{document}                                                        
\renewcommand{\refname}{\normalsize\bf References}
\title{%
Transport Properties and Density of States of Quantum 
Wires with Off-diagonal Disorder
}

\author{%
        P.\ W.\ Brouwer%
        \address{Lyman Laboratory of Physics,
                 Harvard University, Cambridge, MA 02138, USA}%
        \address{Laboratory of Atomic and Solid State Physics,
                 Cornell University, Ithaca, NY 14853, USA}%
        \thanks{Work at Harvard is supported by NSF grants nos. 
        DMR 94-16910, DMR 96-30064, and DMR 97-14725.}
        ,
        \,Christopher Mudry$^{\rm a}$%       
        \address{Paul Scherrer Institut, CH-5232, Villigen PSI,
                 Switzerland}%
\thanks{Supported in part by a grant from the Swiss Nationalfonds.}
        \,and Akira Furusaki%
        \address{Yukawa Institute for Theoretical Physics,
                 Kyoto University, Kyoto 606-8502, Japan}%
\thanks{Supported in part
by Grant-in-Aid for Scientific Research (No.\ 11740199) from the
Ministry of Education, Science, Sports and Culture, Japan.
The numerical calculations were performed at the Yukawa
Institute Computer Facility.}
}
%%%%%%%%%%%%%%%%%%%%%%%%%%%%%%%%%%%%%%%%%%%%%%%%%%%%%%%%%%%%%%%%%%%%%%%%%%
%
% abstract   
%
%%%%%%%%%%%%%%%%%%%%%%%%%%%%%%%%%%%%%%%%%%%%%%%%%%%%%%%%%%%%%%%%%%%%%%%%%%
\begin{abstract}
\hrule
\mbox{}\\[-0.2cm]

\noindent{\bf Abstract}\\

We review recent work on the random hopping problem in a
quasi-one-dimensional geometry of $N$ coupled chains (quantum wire
with off-diagonal disorder).  Both density of states and conductance
show a remarkable dependence on the parity of $N$.  The theory is
compared to numerical simulations.\\[0.2cm]
\hrule
\end{abstract}

\maketitle

\setcounter{footnote}{0}

\section{Introduction and Results}

With the realization that the problem of Anderson localization is
amenable to a RG analysis came the understanding that some transport
and spectral properties of a quantum particle subjected to a weak
random potential depend qualitatively only on dimensionality and the
symmetries of the random potential \cite{Lee85}.  For essentially all
types of randomness the dimensionality two of space plays the role of
the lower critical dimension: Below two dimensions and for arbitrary
weak disorder a quantum particle is always localized (i.e., the
wavefunction is insensitive to a change in boundary conditions, or,
equivalently, exponential decay of the conductance with system size),
whereas a finite amount of disorder is needed to localize a quantum
particle in three dimensions.

A notorious exception to this rule is the case of a particle on
a single chain with random nearest neighbor hopping which appears
in many reincarnations, 
e.g., random $XY$
spin chains \cite{McCoyWu68} and diffusion in random
environments \cite{random diffusion84}.
In all these cases, the cause underlying the
different behavior of random systems with off-diagonal disorder is the 
existence of a sublattice, or {\em chiral}, symmetry, which is absent
for diagonal disorder \cite{chiral always99}.

The purpose of this contribution is to review our recent work 
(together with Simons and Altland \cite{Brouwer98}) on the random 
hopping problem in a quasi-one-dimensional ``wire'' geometry 
\cite{Brouwer98,Mudry99,Brouwer99NON,Brouwer99DOS}.
We consider the density of states (DoS) and conductance, and
focus on the differences between the cases of off-diagonal
and diagonal disorder in the localized regime. (For off-diagonal
disorder, we assume absence of diagonal disorder.) The quantum
wire is the logical intermediate between one and two 
dimensions.\footnote{Two dimensional models with off-diagonal disorder
are also studied in the context of strongly interacting electron systems
\cite{Interactions}. We refer the reader to Ref.\ \cite{Furusaki99} 
for a review of work done on the two dimensional case.}
The quasi-one-dimensional geometry allows us to find exact solutions for 
the conductance at the band center $\varepsilon=0$ and the DoS
near $\varepsilon=0$, while, in contrast to purely
one-dimensional counterparts, it still shows a crossover from a
diffusive to a localized regime, as is the case in two dimensions. 

In the diffusive regime, differences between off-diagonal and diagonal 
disorder are limited to small quantum corrections to the DoS and
conductance. 
The DoS near $\varepsilon=0$ follows from the so-called ``chiral''
random 
matrix theory (RMT), which was originally proposed in the context of
quenched 
approximations to the QCD Hamiltonian \cite{chRMT}. In chiral RMT, 
all eigenvalues come in pairs $\pm \varepsilon$. For $N$
coupled random hopping chains, level
repulsion between the smallest eigenvalue above the band center
$\varepsilon=0$ and its mirror image then leads to a
suppression of the DoS near $\varepsilon=0$,
\begin{equation}
  {d {\cal N}/d \varepsilon} \propto v_F^{-1} 
  (\varepsilon/\Delta)^{\beta-1},
  \ \ 0 < \varepsilon \ll \Delta \label{eq:DOSRMT}.
\end{equation}
Here $d {\cal N}/d \varepsilon$ is the DoS per unit volume, 
$v_F$ is the Fermi velocity, $\Delta=2 \pi v_F/NL$ is the 
mean level spacing, and the symmetry index
 $\beta=1$ in the presence of both time-reversal and 
spin-rotational symmetry and $\beta=2$ ($4$) if time-reversal symmetry
(spin-rotational symmetry) is broken.
Several level spacings away from the center of the
band, the mirror symmetry of the spectrum plays no important role, 
and the level statistics were found to agree with those of the standard
RMT. Similarly, different
quantum corrections to the conductance are obtained for energies close
to the band center \cite{Mudry99}.

On the other hand, beyond the diffusive regime, i.e., for sample lengths 
$L \gtrsim \xi$, where $\xi$ is a crossover length scale (to be defined 
after Eq.\ (\ref{eq:lnGodd}) below), quantum corrections are
dominant, and the differences between diagonal and off-diagonal disorder
can be much more dramatic. While for diagonal disorder, the
DoS is constant as $\varepsilon \to 0$, Dyson showed that for
off-diagonal disorder ${d {\cal N} / d \varepsilon}$ diverges upon
approaching the 
band center 
\cite{Dyson53} :
\begin{equation}
{d {\cal N} \over d \varepsilon} \propto
{1 \over \xi |\varepsilon\ln^3(\varepsilon \xi/v_F)|},
\quad 0 < \varepsilon \xi/v_F \ll 1.
\label{eq:dyson}
\end{equation} 
Note that, in contrast to
the diffusive regime (\ref{eq:DOSRMT}),
the energy scale that governs the divergence
does not depend on the system size, nor does the form of the singularity
depend on the symmetry index $\beta$. 
For $N$ coupled chains, Dyson's result (\ref{eq:dyson}), remains 
valid for odd $N$,\footnote{The proportionality constants in Eqs.\ 
(\ref{eq:dyson}) and (\ref{eq:mean dos if even N}) may depend on $N$.}
but not for even
$N$ \cite{Brouwer99DOS}: For an even number of chains, the DoS diverges 
only logarithmically for $\beta=1$, whereas ${d {\cal N}/ d
\varepsilon}$ shows
a pseudogap for $\beta=2,4$, 
\begin{equation}
{d {\cal N} \over d \varepsilon}\propto {1 \over v_F}
  \left({ \varepsilon \xi \over v_F}\right)^{\beta-1}\left|\ln
{\varepsilon \xi \over v_F} \right|,
\quad 0 < {\varepsilon \xi \over v_F} \ll 1.
\label{eq:mean dos if even N}
\end{equation}
The relevant energy scale, however, remains the same as in Eq.\ 
(\ref{eq:dyson}). 
The even-odd effect in the DoS around $\varepsilon=0$ is accompanied
by an even-odd effect for the conductance $g$ at the band center
\cite{Brouwer98,Mudry99,Brouwer99NON}. For odd $N$, there is 
no exponential localization; $g$ has
a broad distribution, with fluctuations that are of the same order as
the average,
\begin{equation}
\langle \ln g \rangle = 
  -4\sqrt{{L \over \beta \pi \xi}},\ \
\mbox{var}\, \ln g = 
{8 (\pi-2) L\over \beta \pi \xi},
\label{eq:lnGodd}
\end{equation}
where the crossover length scale 
$\xi = (\beta N+2-\beta)\ell/\beta$,
and $\ell$ is the mean free path.
(For small $N$ there are corrections due to the appearance of a
second dimensionless parameter needed to characterize the
off-diagonal disorder, see Ref.\ \cite{Brouwer99NON} for details.)
In contrast, for even $N$, $g$ is exponentially small, its distribution
being close to log-normal,
\begin{equation}
\langle \ln g \rangle =
-{2 L \over \xi} +  \sqrt{8 L \over \beta \pi \xi},\ \ 
\mbox{var}\, \ln g = {8(\pi - 1)L\over \beta \pi \xi}.
\label{eq:lnG}
\end{equation}
In this case, $\xi$ can be interpreted as the localization length.
Away from the band center and for diagonal disorder,
$\langle \ln g \rangle = -(1/2) \mbox{var}\, \ln g = -2L/\xi'$, 
with $\xi' = (\beta N + 2 - \beta) \ell$, irrespective
of the parity of $N$, as for quantum wires with diagonal
disorder \cite{BeenakkerReview}. Note that for even $N$, 
despite the similarities, differences between off-diagonal and
diagonal disorder persist, though they are more subtle.

In the remaining sections of this paper, we review the theory describing
these two parity effects. 

\section{Two microscopic models}

\subsection{Lattice model}

The random hopping model is a lattice model with nearest neighbor
hopping only, where the hopping amplitude takes a different (random)
value 
between adjacent sites. In our case we consider a two-dimensional
square lattice, for which the Schroedinger Equation takes the form
\begin{equation}
  {\cal H} \psi_{i,j} = 
\sum_{\pm} 
\left(
t_{i,j;i\pm1,j} \psi_{i\pm1,j}+ 
t_{i,j;i,j\pm1} \psi_{i,j\pm1}
\right). 
\label{eq:nn rhp}
\end{equation}
Hermiticity requires $t_{i,j;i\pm1,j} = (t_{i\pm1,j;i,j})^{*}$ and
$t_{i,j;i,j\pm1} = (t_{i,j\pm1;i,j})^{*}$. We consider a
quasi-one-dimensional
geometry (length $L$ of the disordered region much larger than its width 
$Na$, where $N$ is the number of chains and $a$ the lattice constant),
with
open boundary conditions in the transverse direction: $t_{0,j;1,j} =
t_{N,j;N+1,j} = 0$. On the left ($j < 0$) and right ($j > L/a$), the 
disordered region is attached to ideal leads 
($t_{i,j;i\pm 1,j} = t_{\|}$, $t_{i,j;i,j\pm1} = t_{\bot}$ for
$j  < 0$ or $j > L/a$). In the disordered region, the hopping amplitudes
show fluctuations around the average values $t_{\|}$ and $t_{\bot}$, 
respectively. The fluctuations are real, 
complex, or quaternion, for the symmetry classes $\beta=1$, $2$, and
$4$, respectively.

The random flux model is a special version of the random hopping model
for which each plaquette is threaded by a random flux. 
In that case, the hopping amplitudes have magnitude $t$ and a
random phase $\phi_{i,j;i\pm1,j}$, so that the total phase accumulated
going around a plaquette equals the phase through the plaquette in units
of the flux quantum $\Phi_0$.

The random hopping model is special because of the existence of a
sublattice
symmetry: Under a transformation 
$$
  \psi_{i,j} \to (-1)^{i+j} \psi_{i,j},
$$
the Hamiltonian changes sign. Hence, if $\varepsilon$ is an
eigenvalue of ${\cal H}$, then $-\varepsilon$ is so as well. The band
center $\varepsilon=0$ is a special energy with respect to this
transformation. As we shall see below, it is the existence of this
extra symmetry that causes the spectral and transport properties of
the random hopping model at the band center to be so dramatically
different from those of models with on-site disorder. For historical
reasons, the sublattice symmetry is referred to as {\em chiral}
symmetry.

\subsection{Continuum model}

While the lattice version (\ref{eq:nn rhp}) of the random hopping model 
is the version that is usually considered in the literature, we use a
different (continuum) model for our analytical calculations,
\begin{equation}
{\cal H}_{\rm cont} \psi (y) =
\left[
{\rm i}\sigma_3\partial_y+
\sigma_3v(y)+
\sigma_2w(y)
\right]\psi(y).
\label{eq:Schr eq cont}
\end{equation}
Here $\psi$ is a $2N$ component vector (elements of $\psi$ occur in
pairs
that correspond to left and right movers), $v$ and $w$ are $N \times N$
random Hermitean matrices, and the $\sigma_\mu$ $(\mu = 1,2,3)$ are the 
Pauli matrices. In Eq.\ (\ref{eq:Schr eq cont}) and below we choose our
units such that the Fermi velocity $v_F = 1$.
The chiral symmetry is now represented by
$$
  \sigma_1{\cal H}_{\rm cont}\sigma_1=-{\cal H}_{\rm cont}.
$$
For $\beta=1$, $w$ ($v$) is (anti-)symmetric; for $\beta=4$, $w$ ($v$)
is (anti)-self dual. The disorder potentials $v$ and $w$ are chosen
independently and Gaussian distributed with zero mean and with variance
$$
\langle
v^{\ }_{ij}(y)v^{\dag}_{kl}(y')
\rangle =
 {{\delta(y-y')} \over\xi} 
\left( \delta^{\ }_{ik}\delta^{\ }_{jl}
-
\mbox{${2-\beta\over\beta}$}
\delta^{\ }_{il}\delta^{\ }_{jk} 
\right),
$$
$$
\langle
w^{\ }_{ij}(y)w^{\dag}_{kl}(y')
\rangle =
{{\delta(y-y')} \over\xi} 
\left( \delta^{\ }_{ik}\delta^{\ }_{jl}
+
\mbox{${2-\beta\over\beta}$}
\delta^{\ }_{il}\delta^{\ }_{jk} 
\right),
$$
respectively. (For small $N$ a slightly more general form of the
variance
is needed, see \cite{Brouwer99NON}.)
The justification for this change of models is our anticipation that, 
for weak disorder and for a sufficiently large system size, only the
fundamental symmetries of the microscopic model are relevant. In order
to verify this assumption of universality, we compare the theoretical
predictions based on Eq.\ (\ref{eq:Schr eq cont}) below to numerical 
simulations of the lattice model (\ref{eq:nn rhp}).

\section{Fokker-Planck approach }

\subsection{General method}

Spectral and transport properties of the continuum model (\ref{eq:Schr
eq cont}) can be studied in a unified way from the
statistical distribution of the {\em reflection matrix}
$r(\varepsilon)$ of the disordered wire. The reflection matrix is
defined from the relation between the amplitudes
$\psi_{\varepsilon}^{\rm iL}$ and $\psi_{\varepsilon}^{\rm oL}$ of an
incoming and an outgoing wavefunction at energy
$\varepsilon$ on the same (left) side of the disordered region 
(Fig.\ 1),
\begin{equation}
  \psi_{\varepsilon}^{\rm oL} = r(\varepsilon) \psi_{\varepsilon}^{\rm
iL}.
\end{equation}
Transport properties (the conductance $G$) follow from the eigenvalues
of
$r^{\dagger} r$, when 
the right side of the quantum wire is attached to a lead (Fig.\ 1a),
\begin{equation}
G = 
{2 e^2\over h}\, g,\ \ g = \mbox{tr}\, 
\left(1-r^{\dagger} r \right), 
\label{eq:Landauer}
\end{equation}
while the DoS is obtained from the eigenphases of
$r$ when the quantum wire is closed at the right side \cite{Schmidt57}
(Fig.\ 1b)
\begin{equation}
  {d{\cal N}\over d\varepsilon} = {1\over2\pi{\rm i} N L}
{\rm tr}
\left(
r^{\dag} {\partial\over\partial\varepsilon} r
\right). \label{eq:dos}
\end{equation}

The distribution of $r$ is computed using a RG
approach: One calculates how the eigenvalues of $r^{\dag}r$ (in the
case of the conductance) or the eigenphases of $r$ (in the case of the
DoS) change when a thin slice is added to the disordered region
(Fig.\ \ref{fig:composition law and RG step}).
If we place the thin slice to the left of the disordered
region, the change of the reflection matrix is given by
\begin{equation}
r^{\ } \to
r^{\ }_1+t'_{1}\left({1}-r^{\ } r'_1\right)^{-1} r^{\ } t^{\ }_1,
\label{eq:r12}
\end{equation}
where $r_1$ and $r_1'$ ($t_1$ and $t_1'$) are the reflection
(transmission)
matrices of the thin slice 
(Fig.\ \ref{fig:composition law and RG step}). 
If the length $\delta L$ of the thin slice is much smaller than a mean
free path, these can be computed using the Born approximation,
\begin{eqnarray}
r^{\ }_{1}&=&
-W+\mbox{${{\rm i}\over2}$}(VW-WV), \nonumber
\\
t^{\ }_{1}&=&
{1}+
{\rm i} V -
\mbox{${1\over2}$}(V^2 + W^2) + {\rm i} \varepsilon \delta L,
\label{eq: t slice} \nonumber
\\
r'_{1}&=&
W+\mbox{${{\rm i}\over2}$}(VW-WV), \nonumber
\\
t'_{1}&=&
{1} -
{\rm i} V -
\mbox{${1\over2}$}(V^2+W^2) + {\rm i} \varepsilon \delta L,
\label{t' slice}
\end{eqnarray}
where
$
V=\int_0^{\delta L}dy\, v(y),
$
and
$
W=\int_0^{\delta L}dy\, w(y).
$
Here we neglected terms that are of order $\delta L^2$.
Together with the distribution of the disorder potentials $w$ and $v$,
Eqs.\ (\ref{eq:r12})--(\ref{t' slice}) define how the distribution
of the reflection matrix $r$ evolves with the length $L$ of the
quantum wire. The RG approach can be represented in terms of
a Fokker-Planck equation describing the ``Brownian motion'' of the 
eigenvalues of $r^{\dag}r$ or eigenphases of $r$
in the geometry of Fig.\ 1a and 1b, respectively. 

Equations (\ref{eq:r12})--(\ref{t' slice}) are exact for the continuum 
model (\ref{eq:Schr eq cont}). A different choice for the distribution 
of $v$ and $w$, or use of the lattice model (\ref{eq:nn rhp}) 
instead of Eq.\ (\ref{eq:Schr eq cont}), would 
have led to different statistical properties of the scattering matrix 
for a thin slice. However, as we have verified numerically, such 
differences are irrelevant in the RG sense, i.e., they disappear for 
sufficiently long  wires (longer than the mean free path $\ell$) and
weak disorder ($\ell$ much larger than the lattice constant $a$).

\subsection{DoS}

Since the quantum wire is closed at one end, the initial condition 
at $L=0$ for the
RG equation (\ref{eq:r12}) is $r=1$. The RG flow ensures that the 
reflection matrix $r$ remains a $N\times N$ unitary matrix for all
$L$. Hence, for all $L$ there exists a Hermitean
$N\times N$ matrix $\Phi$ with $r=\exp({\rm i}\Phi)$. The eigenvalues
$\phi_j$ of $\Phi$ are the eigenphases of $r$. The change of 
$\Phi$ under the increment (\ref{eq:r12})--(\ref{t' slice}) is
$$
\cot\left({\Phi\over2}\right)\rightarrow
e^{-(W+{\rm i}V)}
\cot\left({\Phi\over2}+\varepsilon \delta L \right)
e^{-(W-{\rm i}V)}
$$
up to second order in $V$ and $W$ and first order in $\varepsilon$.
Taking the length $L$ as a fictitious ``time'', the eigenphases 
$\phi_{j}$ of $\Phi$ perform a Brownian motion on the unit circle, which 
is such that upon increasing $L$, they move (on average)
counterclockwise.
Integration of Eq.\ (\ref{eq:dos}) yields a relation between
the $\phi_j$ and the DoS valid in the limit of large $L$,
\begin{equation}
  {\cal N}(\varepsilon') = \int_0^{\varepsilon'} d\varepsilon
    {d {\cal N} \over d\varepsilon}
  = {1 \over 2 \pi N} \sum_{j} {\partial \phi_{j} \over \partial L}.
\end{equation}
[Note that for $\varepsilon=0$, $\phi_{j}=0$ for all $L$.] 
Hence, to find the (average) DoS per unit volume, we compute the 
(average) ``{\em current}'' of the eigenphases $\phi_j$ moving around
the unit circle and differentiate to energy.
We remark that, in the absence of disorder, 
the angles $\phi_{j}$ move around at a constant
speed $\propto \varepsilon$, resulting in a constant DoS.
With disorder, their motion acquires a random (Brownian) component,
which
dramatically affects their average speed, and hence the DoS.

For a quantitative description a different parameterization of 
the eigenphases of the reflection matrix is more convenient, 
\begin{equation}
\tan (\phi_{j}/2) = \exp(u_{j}), \label{eq:phiu}
\end{equation}
where the $u_{j}$ are restricted to the two branches
$\mbox{Im}\, u_{j} = 0$ and $\mbox{Im}\, u_{j} = \pi$ in the
complex plane (Fig.\ \ref{fig:parity for u's}). 
Starting from the RG equation (\ref{eq:r12}), one then obtains
that the joint probability distribution $P(u_1,\ldots,u_N)$
for the $u_{j}$ obeys
\begin{eqnarray}
  {\partial P \over \partial L} &&\hskip -14 pt=
  \sum_{j} {\partial \over \partial u_{j}}
  \left[
   {4 \over \beta \xi} 
  J {\partial\over \partial u_{j}} J^{-1}- 
  2 {\varepsilon} \cosh(u_{j})
  \right] P,
  \nonumber \\
  J &&\hskip -14 pt=
  \prod_{j < k} \sinh^{\beta}[(u_{j} - u_{k})/2].
  \label{eq:FP dos}
\end{eqnarray}
This Fokker-Planck equation describes the motion of $N$ fictitious
Brownian ``particles'' with coordinates $u_{j}$ ($j=1,\ldots,N$) 
and diffusion coefficient $4/\beta \xi$, subject to a driving force
$F_{\varepsilon} = 2\varepsilon \cosh u_j$ and a {\em long-range}
repulsive two-body interaction 
$F_{\rm int} = (2/\xi) \coth [(u_{j} - u_{k})/2]$.
The parameterization (\ref{eq:phiu}) is such that a Brownian 
particle with coordinate $u_{j}$ that vanishes on one of the branches 
at $\pm\infty$ reappears at the opposite branch
(dotted arrows in Fig.\ \ref{fig:parity for u's}). 

Analysis of the Brownian motion described by Eq.\ (\ref{eq:FP dos}) then
leads to the asymptotes (\ref{eq:dyson},\ref{eq:mean dos if even N}) 
for the DoS in the random hopping model. 
A detailed derivation of these results based on an estimate of 
the steady-state-current 
supported by Eq.\ (\ref{eq:FP dos}) can be found in Ref.\
\cite{Brouwer99DOS}. Here we present only
a brief sketch, focusing on the origin of the even-odd effect: The
competition between the driving force $F_{\varepsilon}$ and 
diffusion on the one side,
and the long-range repulsive two-body interaction $F_{\rm int}$ on
the other side. The driving force $F_{\varepsilon}$ 
pushes the particles to the right (left) on the lower (upper)
branch and thus causes the nonzero steady-state-current. Diffusion
keeps the particles mobile where the driving force
$F_{\varepsilon}$ is small, thus enhancing the current, and hence
the DoS \cite{Brouwer99DOS}. 
The long-range repulsion $F_{\rm int}$, on the other hand, 
traps an equal number of the particles in two ``traps'' near 
$u = \ln(\varepsilon \xi)$ and $u = -\ln(\varepsilon \xi)
 + \pi i$, and 
thus tends to suppress the DoS. For even $N$, all particles are trapped 
($N/2$ particles in each trap). Rare events caused by particles that are 
``thermally'' excited out of 
their traps still allows for a small ``current'', resulting in
the small DoS in Eq.\ (\ref{eq:mean dos if even N})
(Fig.\ \ref{fig:parity for u's}a).
For odd $N$, however, only $(N-1)/2$ particles are trapped near each 
end point, while one particle is left alone, unaffected by the
long-range repulsion (Fig.\ \ref{fig:parity for u's}b). 
Motion of this particle is dominated by diffusion, which
explains the enhancement of the DoS for odd $N$.

\begin{figure}
\epsfxsize=0.8\hsize
\hspace{0.03\hsize}
\epsffile{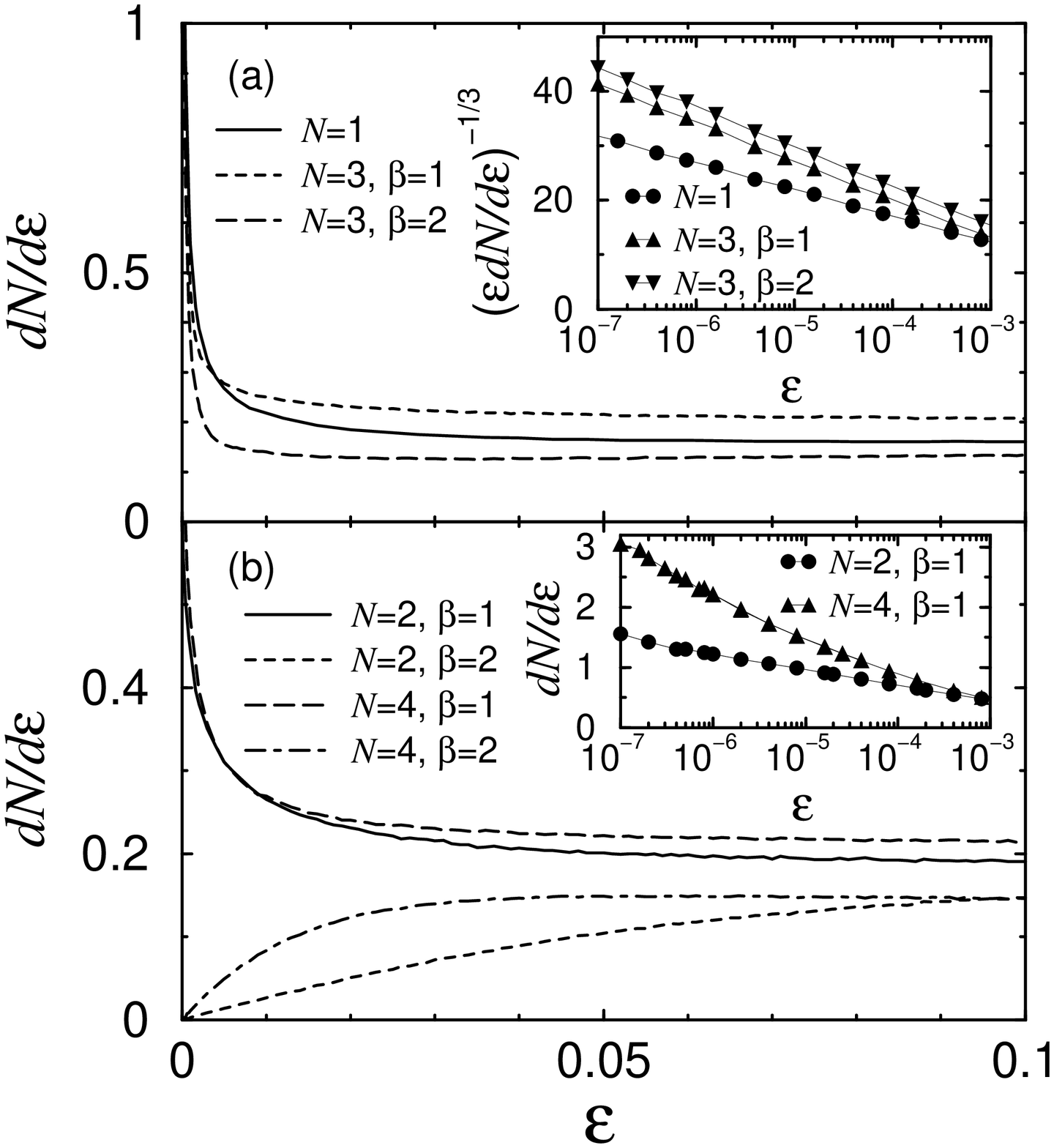}
\refstepcounter{figure}
%\vskip -1cm

FIG.\ \ref{fig:DoS}.
%\caption
\small{
Density of states, from numerical simulations for $N=1,3$ (a) 
and for $N=2,4$ (b).  The data shown on the linear scale are computed
for $L=200$ ($L=500$ for $N=1$) while the data in the insets are for
$L=10^5$ ($L=10^6$ for $N=1$). For $\beta=1$ and also for $N=1$ the
hopping amplitudes 
are taken from a uniform distribution in the interval [0.5,1.5], while
for
$\beta=2$ the random flux model \cite{Mudry99} is used, where
the randomness is introduced only via the random phases of the
hopping amplitudes. Results of an average over $4\times10^4$--$10^6$
disorder realizations are shown. 
}\hfill
\label{fig:DoS}
\end{figure}

The agreement between the theory and numerical simulation of the lattice
random hopping model is excellent
as is depicted in Fig.\ \ref{fig:DoS}.

\subsection{Conductance}

The chiral symmetry results in the symmetry relation 
$r(\varepsilon)=r(-\varepsilon)^{\dag}$ \cite{Mudry99}.
Hence, at the band center, the reflection matrix is Hermitean. We
parameterize the
eigenvalues of $r$ as 
$\tanh x_j$, where the $x_j$ are $N$ real numbers, and 
formulate the RG approach of Sec.\ 3.1 as a Brownian motion process
for the $x_j$.
Away from the band center, only the eigenvalues $\tanh^2 x_j$ of the
product
$r^{\dag}r$ can be studied in the RG approach: In this case, the sign of 
the $x_j$ has no relevance, since $r^{\dagger} r$ does not change 
under $x_j\rightarrow-x_j$.

In terms of the $x_j$, the
conductance reads
\begin{equation}
  g =  \sum_{j} (1-\tanh^2 x_j) 
  =  \sum_{j} \cosh^{-2} x_j.
\end{equation}
The RG equation (\ref{eq:r12}) then leads to a Fokker-Planck equation
for the $L$-evolution of the joint probability distribution
$P(x_1,\ldots,x_N)$
at the band center
\cite{Brouwer98,Mudry99,Brouwer99NON},
\begin{eqnarray}
&&
{\partial P\over\partial L}=
{1 \over \beta \xi}
\sum_{j=1}^N
{\partial\over\partial x_j} 
\bigg[
J{\partial\over\partial x_j} \left(J^{-1}P\right)
\bigg],
\label{eq:Jch}\\ &&
J = \prod_{k>j}|\sinh(x_j-x_k)|^\beta.
\nonumber
\end{eqnarray}
Sufficiently far away from the band center, the $L$-evolution of
the $x_j$'s is the same as for a quantum wire with on-site
disorder, which is given by the well-known DMPK equation \cite{DMPK}
\begin{eqnarray}
&&\hskip-4pt
{\partial P\over\partial L}=
{1 \over 2 \xi'}
\sum_{j=1}^N
{\partial\over\partial x_j} 
\left[
J{\partial\over\partial x_j} \left(J^{-1}P\right)
\right],
\label{eq:JDMPK}\\
&&\hskip - 4pt
J=
\prod_{k>j}|\sinh^2 x_j-\sinh^2 x_k|^\beta
\prod_k|\sinh(2x_j)|,
\nonumber
\end{eqnarray}
where $\xi' = (\beta N + 2 - \beta) \ell$.

Both Fokker-Planck equations describe
the Brownian motion of $N$ particles with the coordinate  
$x_j$ on the real axis and interacting through a hardcore repulsive 
two-body potential (Fig.\ \ref{fig:parity for x's}). 
Far away from the band center, 
a particle at position $x_j$ is also repelled by the
mirror image at position $-x_j$ as is required by the symmetry
of $r^{\dag}r$ under $x_j\rightarrow-x_j$. For each $x_j$, the
repulsion from the mirror image $-x_j$ ensures that
all $x_j$ increase linearly with $L$, thus causing the
exponential decay of $g$ with $L$. 
On the other hand, at the band center the sign of $x_j$ matters
since
there is no repulsion from mirror images. For odd $N$
there will be no net force on the particle with coordinate $x_{(N+1)/2}$
(for large $L$). Hence $x_{(N+1)/2}$ remains close to the origin,
while all other $x_j$ increase (or decrease) linearly with $L$. 
The coordinate $x_{(N+1)/2}$ dominates the conductance, its
fluctuations leading to the 
$(L/\xi)^{1/2}$ dependence of $\ln g$ in Eq.\ (\ref{eq:lnGodd}).
For even $N$ no such ``exceptional''
particle exists; all $x_j$ increase (if they are positive) or
decrease (if the are negative) linearly with $L$, causing the
conductance to be exponentially small, cf.\ Eq.\ (\ref{eq:lnG}).

\begin{figure}
\epsfxsize=0.8\hsize
\hspace{0.03\hsize}
\epsffile{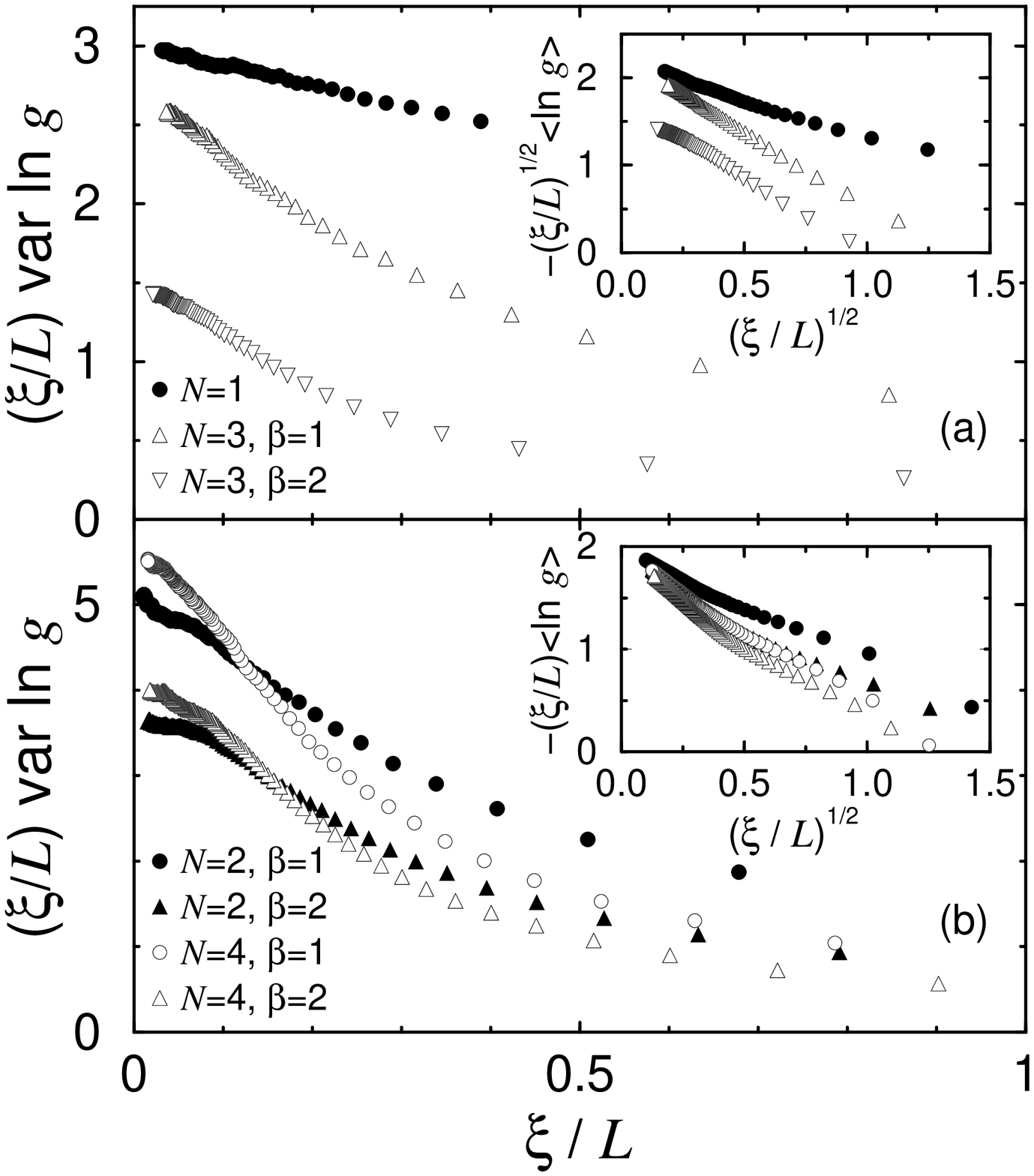}
\refstepcounter{figure}
%\vskip -4mm

FIG.\ \ref{fig:ln g}.
%\caption
\small{Average and variance of $\ln g$ versus $L/\xi$ for $N=1,3$ (a) 
and $N=2,4$ (b). The hopping amplitude $t_{\|}$ is taken from a uniform
distribution in the interval [0.9,1.1], while $t_{\bot}$ is a 
real random numbers in [-0.2,0.2] for $\beta=1$ and a
complex random numbers with modulus $< 0.1$ for $\beta=2$. The
characteristic length $\xi$ is obtained from a comparison with 
Eqs.\ (\protect\ref{eq:lnGodd}) and  (\protect\ref{eq:lnG}).
%The characteristic length $\xi=623a$ for $N=1$, and $\xi=407a$,
%$633a$, $508a$, $518a$, $629a$, and $1083a$ for $(N,\beta)=(2,1)$,
%(2,2), (3,1), (3,2), (4,1), and (4,2), respectively.
Results of an average over more than $2\times10^4$ disorder
realizations are shown.
}\hfill
\label{fig:ln g}
\end{figure}

We refer the reader to Refs.\ \cite{Mudry99}
for the detailed calculation of the moments of the conductance.
The agreement between the theory and numerical simulations of 
the random flux case is excellent as is illustrated 
in Fig.\ \ref{fig:ln g}.

\newpage

\begin{figure}
\begin{center}
\begin{picture}(200, 30)(0,10)
% (A,B)(C,D) specifies a box of width A, height B with a lower-left
%corner
% located at (C,D). By default, (C,D)=(0,0).
%%%%%%%%%%%%%%%%%%%%%%%%%%%%%%%%%%%%%%%%%%%%%%%%%%%%%%%%%%%%%%%%%%%%%%%
\put(  0,12){({\rm a})}            \put(110,12){({\rm b})}
\thicklines
\put( 20, 0){\line(1,0){80}} \put(130, 0){\line(1,0){80}}
\put( 20,20){\line(1,0){80}} \put(130,20){\line(1,0){80}}
\put( 20, 5){$ \leftarrow$}  \put(130, 5){$ \leftarrow$}
\put( 20,10){$\rightarrow$}  \put(130,10){$\rightarrow$}
\thicklines
\put( 30, 0){\line(0,1){20}} \put(140, 0){\line(0,1){20}}
\put( 90, 0){\line(0,1){20}} \put(210, 0){\line(0,1){20}}
                             \put(209.5,0){\line(0,1){20}}
                             \put(209  ,0){\line(0,1){20}}
\thinlines
\multiput( 30, 20)(5,0){9}{\line(1,-1){20}}
                             \multiput(140,
20)(5,0){11}{\line(1,-1){20}}
\put( 30, 15){\line(1,-1){15}}
\put( 30, 10){\line(1,-1){10}}
                             \put(140, 15){\line(1,-1){15}}
                             \put(140, 10){\line(1,-1){10}}
\put( 90,  5){\line(-1,1){15}}
\put( 90, 10){\line(-1,1){10}}
                             \put(210,  5){\line(-1,1){15}}
                             \put(210, 10){\line(-1,1){10}}

\end{picture}
\hfill\break
\end{center}
FIG.\ \ref{fig:quantum wires with BC}.
\refstepcounter{figure}
%\caption
\small{
Boundary conditions used for the computation of the 
conductance (a) and DoS (b). 
%
%In (a) the reflection matrix for
%reflection from the left is computed with an ideal lead attached
%to the right of the quantum wire; In (b) the wire is closed on
%the right side by a hard wall.
%
%Quantum wire with a disordered (hashed) region of length $L$
%for two different boundary conditions:
%(a) Ideal leads are attached to both sides of the disordered region
%    for the conductance.
%(b) Ideal lead is attached to the left of the disordered region
%    with a hard wall on the right for the DoS.
}
\hfill
\label{fig:quantum wires with BC}
\end{figure}

\begin{figure}
\begin{center}
\begin{picture}(200,40)(13,-170) 
% (A,B)(C,D) specifies a box of width A, height B 
% with a lower-left corner
% located at (C,D). By default, (C,D)=(0,0).
%
\thicklines
\put( 20,-120){\line(1,0){200}}
\put( 20,-160){\line(1,0){200}}
%
% Horizontal lines defining the wire
%
\multiput( 44,-160)(12,0){2}{\line(0,1){40}} 
\multiput(120,-160)(40,0){1}{\line(0,1){40}} 
\thinlines
%
% 4 vertical lines defining disordered regions 1 and 2
%
\multiput(44,-120)(0,-4){8}{\line(1,-1){12}} 
\put( 44,-180){$\delta L$}
\put(46,-168){\vector( 1,0){10}}
\put(54,-168){\vector(-1,0){10}}
%
% Hash marking disorder for region 1
%
\multiput(208,-120)(-4, 0){23}{\line( 1,-1){12}} 
\multiput(208,-148)(-4, 0){23}{\line( 1,-1){12}} 
\multiput(120,-120)( 0,-4){ 7}{\line( 1,-1){12}}
\multiput(208,-120)( 0,-4){ 7}{\line( 1,-1){12}}
\multiput(196,-120)(-4, 0){20}{\line( 1,-1){24}} 
\multiput(196,-136)(-4, 0){20}{\line( 1,-1){24}} 
%\multiput(120,-120)( 0,-4){ 4}{\line( 1,-1){24}}
%\multiput(208,-120)( 0,-4){ 4}{\line( 1,-1){24}}
%
% Hash marking disorder for region 2 
%
\put( 20,-133){$t'    _1\leftarrow    $}
\put( 20,-150){$r^{\ }_1\hookleftarrow$}
\put( 58,-133){$\rightarrow     t^{\ }_1$}
\put( 58,-150){$\hookrightarrow r'    _1$}
%\put( 96,-133){$t'    _2\leftarrow    $}
\put( 96,-150){$r        \hookleftarrow$}
%\put( 90,-150){$\hphantom{r^{\ }_2}\hookleftarrow$}
%
% Putting arrows, reflection and transmission matrices
%
%\put(164,-142){$r_2$}
\put(164,-180){$L$}
\put(220,-168){\vector(-1,0){100}}
\put(120,-168){\vector( 1,0){100}}
%
% Putting S matrices in center disordered regions
%
\end{picture}
\hfill\break
\end{center}
FIG.\ \ref{fig:composition law and RG step}.
\refstepcounter{figure}
%\caption
\small{
A thin disordered slice of length $\delta L$
is added to the left of a disordered wire
of length $L$.
}
\label{fig:composition law and RG step}
\end{figure}

\begin{figure}
\begin{center}
\begin{picture}(200,100)( 0, 0)
% (A,B)(C,D) specifies a box of width A, height B with a lower-left
%corner
% located at (C,D). By default, (C,D)=(0,0).

\put(  0,100){$({\rm a})$}
\linethickness{0.2mm}
\put( 30,100){\line(1,0){160}}
\put( 30, 60){\line(1,0){160}}
%
% Defines two branches
%
\qbezier[30]( 30,100)( 10, 80)( 30, 60)
\qbezier[30](190, 60)(210, 80)(190,100)
%
% Makes a curve by interpolating a parabola through two end points
% and one reference point: 
% \qbezier[ # points] (initial, coo) (ref, coo) (final, coo)
%
\put( 25, 65){\vector (1,-1){5}}
\put(195, 95){\vector (-1,1){5}}
%
% Adds arrows on curves
%
\multiput( 35, 60)( 6, 0){5}{\circle*{4}}
%\put( 60, 66){$\longleftarrow$}
\multiput(185,100)(-6, 0){5}{\circle*{4}}
%\put(145, 90){$\longrightarrow$}
%%%%%%%%%%%%%%%%%%%%%%%%%%%%%%%%%%%%%%%%%%%%%%%%%%%%%%%%%%%%%%%%%%%%%%%%%%%
\put(  0, 40){$({\rm b})$}
\linethickness{0.2mm}
\put( 30, 40){\line(1,0){160}}
\put( 30,  0){\line(1,0){160}}
%
% Defines two branches
%
\qbezier[30]( 30, 40)( 10, 20)( 30,  0)
\qbezier[30](190,  0)(210, 20)(190, 40)
%
% Makes a curve by interpolating a parabola through two end points
% and one reference point: 
% \qbezier[ # points] (initial, coo) (ref, coo) (final, coo)
%
\put( 25,  5){\vector (1,-1){5}}
\put(195, 35){\vector (-1,1){5}}
%
% Adds arrows on curves
%
\multiput( 35,  0)( 6, 0){5}{\circle*{4}}
%\put( 60,  6){$\longleftarrow$}
\multiput(185, 40)(-6, 0){5}{\circle*{4}}
%\put(145, 30){$\longrightarrow$}
\put(110,  0){\circle*{4}}
\end{picture}
\end{center}
FIG.\ \ref{fig:parity for u's}.
\refstepcounter{figure}
%\caption
\small{
The $L$-dependence of the eigenphases of $r$ is described in terms
of a Brownian motion of fictitious particles with coordinate $u_j$ 
where $\mbox{Im}\, u = 0$ or $\pi$. If $N$ is even, 
a repulsive interaction (see text) traps all particles near 
$u = -\ln \varepsilon \xi$ or $\ln \varepsilon \xi + {\rm i} \pi$ 
and prohibits
them from moving around (a). If $N$ is odd, one particle can diffuse
freely around the branches, and thus leads to a significantly higher
DoS than for even $N$ (b).}\hfill
\label{fig:parity for u's}
\end{figure}

\begin{figure}
\begin{center}
\begin{picture}(200,100)( 0,10)
% (A,B)(C,D) specifies a box of width A, height B with a lower-left
%corner
% located at (C,D). By default, (C,D)=(0,0).
\thicklines
\multiput(20,80)(0,-30){2}{$\line(1,0){180}$}
\put(110,20){$\line(1,0){90}$}
\put(93,20){$\line(1,0){17}$}
\put(53,20){$\line(1,0){34}$}
\put(20,20){$\line(1,0){27}$}
\thinlines
\multiput(110,44)(0,5){10}{$\line(0,1){2}$}
\multiput(110, 5)(0,5){7}{$\line(0,1){2}$}
%%%%%%%%%%%%%%%%%%%%%%%%%%%%%%%%%%%%%%%%%%%%%%%%%%%%%%%%%%%%%%%%%
\put(0,78){$({\rm a})$}
\multiput( 50,80)(40,0){4}{{\circle*{6}}}
\put( 40,70){$x_{{N\over2}+3}$}\put( 60,88){\vector(-1,0){30}}
\put( 80,70){$x_{{N\over2}+1}$}\put(100,88){\vector(-1,0){15}}
\put(120,70){$x_{{N\over2}  }$}\put(120,88){\vector( 1,0){15}}
\put(160,70){$x_{{N\over2}-1}$}\put(160,88){\vector( 1,0){30}}
%%%%%%%%%%%%%%%%%%%%%%%%%%%%%%%%%%%%%%%%%%%%%%%%%%%%%%%%%%%%%%%%%
\put(0,48){$({\rm b})$}
\multiput( 50,50)(30,0){5}{{\circle*{6}}}
\put( 40,40){$x_{{N+5\over2}}$}\put( 60,58){\vector(-1,0){30}}
\put( 70,40){$x_{{N+3\over2}}$}\put( 90,58){\vector(-1,0){15}}
\put(100,40){$x_{{N+1\over2}}$}
\put(130,40){$x_{{N-1\over2}}$}\put(130,58){\vector( 1,0){15}}
\put(160,40){$x_{{N-3\over2}}$}\put(160,58){\vector( 1,0){30}}
%%%%%%%%%%%%%%%%%%%%%%%%%%%%%%%%%%%%%%%%%%%%%%%%%%%%%%%%%%%%%%%%%
\put(0,18){$({\rm c})$}
\multiput( 50,20)(40,0){2}{{\circle{6}}}
\multiput(130,20)(40,0){2}{{\circle*{6}}}
\put( 33,10){$-x_{N-1}$}\put( 60,28){\vector(-1,0){30}}
\put( 73,10){$-x_{N  }$}\put(100,28){\vector(-1,0){15}}
\put(120,10){$ x_{N  }$}\put(120,28){\vector( 1,0){15}}
\put(160,10){$ x_{N-1}$}\put(160,28){\vector( 1,0){30}}
%%%%%%%%%%%%%%%%%%%%%%%%%%%%%%%%%%%%%%%%%%%%%%%%%%%%%%%%%%%%%%%%%
\end{picture}
\end{center}
FIG.\ \ref{fig:parity for x's}.
\refstepcounter{figure}
%\caption
\small{
For a wire with off-diagonal disorder, the $L$-dependence of the 
eigenvalues $\tanh^2 x_j$ of $r^{\dag}r$
is described in terms of a Brownian motion of fictitious particles
 with coordinate $x_j$
on the real axis.
If $N$ is even, all $x_j$ (full circles) repel 
away from 0 (a), while $x_{(N+1)/2}$ 
remains close to 0 for odd $N$ (b). 
In the case of diagonal disorder, all $x_j$ repel from $0$ due to
repulsion from mirror images (open circles) (c).
}
\label{fig:parity for x's}
\end{figure}

\end{document}